\documentclass[sigconf, authorversion]{acmart}

\usepackage{balance}
\usepackage{soul}
\usepackage{xcolor}
\usepackage{tabularx}
\usepackage{subcaption}
\usepackage{enumitem}
\usepackage{graphicx}
\graphicspath{{figs/}}

\newcommand{\blinduni}[1]{#1}
\newcommand{\blind}[1]{#1}

\AtBeginDocument{%
  \providecommand\BibTeX{{%
    \normalfont B\kern-0.5em{\scshape i\kern-0.25em b}\kern-0.8em\TeX}}}

\copyrightyear{2020} 
\acmYear{2020} 
\setcopyright{acmlicensed}
\acmConference[ACMSE 2020]{2020 ACM Southeast Conference}{April 2--4, 2020}{Tampa, FL, USA}
\acmBooktitle{2020 ACM Southeast Conference (ACMSE 2020), April 2--4, 2020, Tampa, FL, USA}
\acmPrice{15.00}
\acmDOI{10.1145/3374135.3385277}
\acmISBN{978-1-4503-7105-6/20/03}

\begin{document}
% \pagestyle{empty}

%% The "title" command has an optional parameter,
%% allowing the author to define a "short title" to be used in page headers.
\title{Toward Predicting Success and Failure in CS2: A Mixed-Method Analysis}

\author{Lucas Layman}
\orcid{0000-0002-2534-8762}
\affiliation{%
  \institution{University of North Carolina, Wilmington}
  \streetaddress{601 S. College Rd.}
  \city{Wilmington}
  \state{North Carolina, USA}
  \postcode{28403}
}
\email{laymanl@uncw.edu}

\author{Yang Song}
% \orcid{0000-0002-2534-8762}
\affiliation{%
  \institution{University of North Carolina, Wilmington}
  \streetaddress{601 S. College Rd.}
  \city{Wilmington}
  \state{North Carolina, USA}
  \postcode{28403}
}
\email{songy@uncw.edu}

\author{Curry Guinn}
% \orcid{0000-0002-2534-8762}
\affiliation{%
  \institution{University of North Carolina, Wilmington}
  \streetaddress{601 S. College Rd.}
  \city{Wilmington}
  \state{North Carolina, USA}
  \postcode{28403}
}
\email{guinnc@uncw.edu}

%%
%% By default, the full list of authors will be used in the page
%% headers. Often, this list is too long, and will overlap
%% other information printed in the page headers. This command allows
%% the author to define a more concise list
%% of authors' names for this purpose.
% \renewcommand{\shortauthors}{Trovato and Tobin, et al.}

%%
%% The abstract is a short summary of the work to be presented in the
%% article.

% SIGCSE 2020: 250 word limit
\begin{abstract}
  Factors driving success and failure in CS1 are the subject of much study but less so for CS2. This paper investigates the \textit{transition from CS1 to CS2} in search of leading indicators of success in CS2. Both CS1 and CS2 at \blinduni{the University of North Carolina Wilmington (UNCW)} are taught in Python with annual enrollments of 300 and 150 respectively. In this paper, we report on the following research questions: 
  \begin{enumerate}
      \item Are CS1 grades indicators of CS2 grades? 
      \item Does a quantitative relationship exist between CS2 course grade and a modified version of the SCS1 concept inventory?
      \item What are the most challenging aspects of CS2, and how well does CS1 prepare students for CS2 from the student's perspective?
  \end{enumerate}
  We provide a quantitative analysis of 2300 CS1 and CS2 course grades from 2013--2019. In Spring 2019, we administered a modified version of the SCS1 concept inventory to 44 students in the first week of CS2. Further, 69 students completed an exit questionnaire at the conclusion of CS2 to gain qualitative student feedback on their challenges in CS2 and on how well CS1 prepared them for CS2.

  We find that 56\% of students' grades were lower in CS2 than CS1, 18\% improved their grades, and 26\% earned the same grade. Of the changes, 62\% were within one grade point. We find a statistically significant correlation between the modified SCS1 score and CS2 grade points. Students identify linked lists and class/object concepts among the most challenging. Student feedback on CS2 challenges and the adequacy of their CS1 preparations identify possible avenues for improving the CS1-CS2 transition.
  
\end{abstract}

%%
%% The code below is generated by the tool at http://dl.acm.org/ccs.cfm.
%% Please copy and paste the code instead of the example below.
%%


%%
%% Keywords. The author(s) should pick words that accurately describe
%% the work being presented. Separate the keywords with commas.

%% This command processes the author and affiliation and title
%% information and builds the first part of the formatted document.
\maketitle

\section{Introduction}

CS1 and CS2 are important first courses for those interested in computing. CS1 is often an introduction to computer science and programming, and CS2 is often an introduction to data structures and algorithms, though the subject matter varies across institutions~\cite{Hertz2010}. Students' performance and experience with the introductory programming sequence have a major impact on retention in the Computer Science major and has been often studied (e.g.,~\cite{Turner2007, Pappas2016}. While students in CS1 may be dipping their toe in the water, students in CS2 have committed to a curricular path and failure in the CS2 course can be more impactful. We focus our research on the \textit{transition from CS1 to CS2}. We want to improve student outcomes in CS2 by both identifying challenges with CS2 topics in particular, and ensuring that CS1 is best preparing students for CS2. 

Both CS1 and CS2 at \blinduni{the University of North Carolina Wilmington (UNCW)} are taught in Python with annual enrollments of 300 and 150 respectively. 
Approximately 15\% of students enrolled in CS2 at \blinduni{UNCW} do not earn the requisite 'C' grade to proceed to the next course in the sequence. Prior research has identified indicators of CS2 final grade, including overall Grade Point Average (GPA), number of absences, and performance in prerequisite courses including CS1~\cite{Enbody2009, Bisgin2018}, but these indicators are not necessarily actionable. 

Our long term research goal is to investigate successful and unsuccessful transitions from CS1 to CS2 to identify interventions and pedagogical improvements in those courses. We begin with the following research questions:
  \begin{enumerate}
      \item Are CS1 grades indicators of CS2 grades? 
      \item Does a quantitative relationship exist between CS2 course grade and a modified version of the Second Computer Science 1 (SCS1) concept inventory~\cite{Parker2016}? The SCS1 evaluates students' understanding of sequencing, selection, iteration, and other CS1 concepts.
      \item What are the most challenging aspects of CS2, and how well does CS1 prepare students for CS2 from the student's perspective?
  \end{enumerate}

For (1), we perform a quantitative analysis of 2300 course grades in CS1 and CS2 at \blinduni{UNCW} from 2013-2019 to evaluate the relationship between CS1 and CS2 grades. 
For (2), we quantitatively evaluate SCS1 assessment~\cite{Parker2016} scores of CS1 skills against student grades for two sections of CS2 offered in Spring 2019.
For (3), we analyze responses to a student questionnaire at the end of CS2 that obtained feedback on challenges faced in the course and on the adequacy of CS1 in preparing them for CS2. We present the most common responses from students and discuss their implications on avenues for pedagogical and curricular improvements.\footnote{This study is approved by \blinduni{UNCW}'s Institutional Review Board: \#19-0129.}

The rest of the paper is organized as follows: Section~\ref{sec:related} discusses prior research on CS2 performance, Section~\ref{sec:curriculum} describes the CS1 and CS2 courses at \blinduni{UNCW}, Sections~\ref{sec:grade_analysis} and \ref{sec:leading_indicators} present quantitative analyses of CS1 vs. CS2 grade and modified SCS1 score vs. CS2 grade, Section~\ref{sec:questionnaire} presents our qualitative questionnaire findings, and we conclude in Section~\ref{sec:conclusion}.

\section{Related Work}
\label{sec:related}

Our CS1 and CS2 courses are taught in Python. Koulouri et al.~\cite{Koulouri2014} report statistical findings that students better learned introductory programming concepts in Python (a "syntactically simple" language) than in Java, however Alzharani et al. found that beginning students struggle with assignments in Python as much or more than in C++~\cite{Alzahrani2018}. Educators have argued that that Python is a good choice for a first programming language~\cite{Radenski2006, Loui2008} because data typing, memory management, and object references are implicit, and that beginning programmers may find such concepts confusing. These concepts are usually important in a CS2 data structures course, though Enbody et al.~\cite{Enbody2009} found no statistical difference in a C++ CS2 data structures course between students who took CS1 in Python vs. C++. 

We are interested in indicators of \textit{unsuccessful} transitions from CS1 to CS2. In this paper, we use \textit{failure to pass CS2} as an indicator of an unsuccessful transition as it can be studied at scale -- more fine-grained measures, such as failing early CS2 homeworks or exams, are warranted in future work. A host of literature reports the many personal~\cite{Bergin2005b,Pappas2016} societal~\cite{Margolis2002}, institutional~\cite{Cohoon2001}, and pedagogical~\cite{Zingaro2015, Vihavainen2014} drivers that lead to student success and failure in CS courses in general. Overall GPA, CS1 grade, and perceived difficulty of prerequisite courses~\cite{Enbody2009, Bisgin2018} have been correlated with CS2 final grades, but these indicators are not actionable. Carter et al.~\cite{Carter2017} used measures of compilation, debugging, and execution in the development environment coupled with the frequency of message board interaction to develop a model that accounted for much of the variance in students' final CS2 grade in a C++ data structures course. Falkner and Falkner~\cite{Falkner2012} find that students who turn in their first assignments on time statistically perform better across the CS curriculum. 

Prior studies~\cite{Sanders2012, Yeomans2019} suggest that common CS2 topics are so-called \textit{threshold concepts}~\cite{Meyer2005} that transform one's understanding of computing and programming, including data structures, classes and inheritance, abstraction, and pointers. In general, research on CS1 concept acquisition greatly outnumbers research on CS2 concept acquisition, though CS2 is receiving increased focus (e.g., ~\cite{Karpierz2014, Paul2013a}). Zingaro et al.~\cite{Zingaro2018} recently presented the most comprehensive study of data structure misconceptions based on qualitative analysis of 279 students' responses to final exam questions. Their study and Porter et al.'s~\cite{Porter2018} report on data structures learning goals led to the recent publication of the Basic Data Structures Inventory (BDSI), a multi-language, validated mechanism for assessing student understanding of linked list, array, binary tree, and binary search tree concepts~\cite{Porter2019}. Unfortunately, the BDSI was published after it could be incorporated into our current study.

Our work expands and adds to this body of literature in several ways. First, our study of the distributions between CS1 and CS2 grade is a large sample size (n=614), providing quantitative evidence on a scale that can overcome biases such as instructor pedagogical effects. Second, no studies have been reported that quantitatively evaluate derivatives of the SCS1 concept inventory~\cite{Parker2016} (discussed in Section~\ref{sec:leading_indicators}) as an indicator of CS2 success. Finally, we ask students to qualitatively reflect on CS2 challenges similar to Zingaro et al.~\cite{Zingaro2018}, but also to consider how well CS1 prepared them for CS2. Educators should try to identify the students who are likely to have unsuccessful transitions from CS1 to CS2 as early as possible. By identifying which factors contribute to failure in CS2, we will be able to focus on how to improve aspects related to both CS1 and CS2 to reduce dropout rates.

\section{Description of CS1 and CS2 courses at \blinduni{UNCW}}
\label{sec:curriculum}

\begin{figure}[tb]
    \centering
    \includegraphics[width=\linewidth]{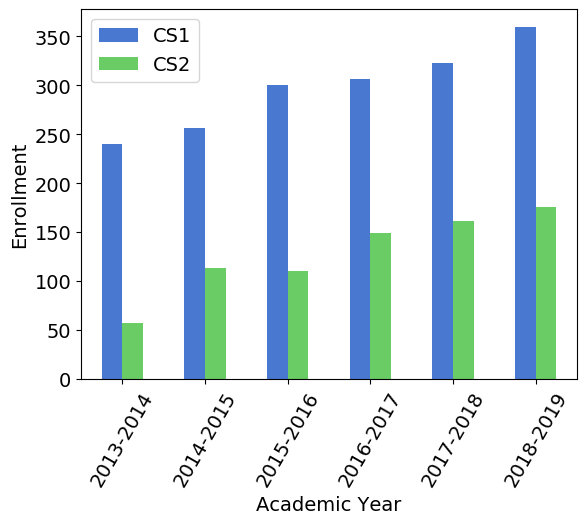}
    \caption{Enrollment History for CS1 and CS2 since Adopting Python}
    \label{fig:enrollment}
\end{figure}

\blinduni{UNCW} is an R2 Doctoral university per the Carnegie Classification. The Department of Computer Science offers undergraduate degree programs in Computer Science (\texttt{\char`\~}420 majors), Information Technology (\texttt{\char`\~}110 majors), and \blind{Digital Arts} (\texttt{\char`\~}60 majors) as well as graduate programs in Computer Science and Information Systems (\texttt{\char`\~}35 students) and Data Science (\texttt{\char`\~}45 students).  Our undergraduate major in Computer Science has received ABET accreditation since 2010-2011. Figure~\ref{fig:enrollment} shows enrollment in CS1 and CS2 each semester. 

% 33 CSC minors
% 59 DA minors
% 5 Data Science minors
% 17 IT minors

In this paper, CS1 and CS2 correspond to the terminology expressed in the ACM Computing Curriculum~\cite{Austing1979} guidelines where CS1 introduces basic algorithm design and programming concepts while CS2 introduces more advanced data abstractions and data structures. Both CS1 and CS2 are four credits (200 minutes per week) equally divided between lecture and computer lab activities. No online sections are offered. All lectures and labs are led by a faculty member who is assisted by a graduate student who provides lab assistance, office hours, and grading help.    

CS1 is \blind{CSC 131 -- Introduction to Computer Science}.  Six to eight sections are offered each semester (1-3 sections in the summer) with approximately 20-25 students per section. The prerequisite for CS1 is a course in college algebra; no programming experience is assumed. CS1 has been taught using the Python programming language since Fall 2013 (Java was the language before that year).  Typically, instructors introduce the IDLE development environment as well as command-line interpretation. The official learning outcomes for CS1 include a demonstrated ability to: understand and implement basic programming concepts (data types, conditionals, function definition), apply problem-solving techniques, use program control structures, practice modular programming, and use file input/output used to solve a variety of problems. The teaching methods and course structure are left to each instructor so long as they address the learning outcomes. All instructors implement a mix of homework and exam, and all courses are taught in a computer lab with programming exercises during most (if not all) sessions.

CS2 is \blind{CSC 231 -- Introduction to Data Structures}.  Three to four sections are offered each semester (one in the summer) with approximately 20-25 students per section. The prerequisite for CS2 is CS1 (grade of C or higher) and Discrete Mathematics which is listed as a pre- or co-requisite. CS2 has been taught using the Python programming language since Spring 2014 (Java was the language before that year).  Typically, instructors introduce a more advanced IDE such as PyCharm or Visual Studio Code.  The official learning outcomes for CS2 include: learning multiple data structures for ordered and unordered data (e.g., lists, trees, hash tables), developing algorithms that implement the main operations of those data structures, performing Big-O analysis, and implementing projects that solve problems using those data structures. Instructors use a variety of in-class programming and written assessments combined with homeworks. Some implementations of the class are taught strictly in a computer lab, while others have two traditional lecture sections and one longer laboratory meeting each week.

Importantly, students must earn a grade of C or higher to enroll in future courses for which CS1 or CS2 is a prerequisite. These courses are required for Computer Science majors and minors as well as \blind{Digital Arts} majors. Students who earn less than a C may repeat a course to earn a higher grade.

\section{Analysis of the relation between CS1 grades and CS2 grades}
\label{sec:grade_analysis}
Our first research question is: \textit{Are CS1 grades indicators of CS2 grades?} We look to validate previous research that found a correlation between CS1 and CS2 grades~\cite{Enbody2009, Bisgin2018}. 
Historical grade distributions also provide context and identify potential biases for interpreting our other analyses. 
\subsection{Grade Point Data Preparation}

We collected all registered course grades for CS1 and CS2 from Fall 2013 (when first offered in Python) through Spring 2019. We mapped the letter grades of [F, D-, D, D+, ..., B, B+, A-, A], to a 4.0 grade point scale. 
We collected 1641 CS1 letter grades and 706 CS2 letter grades in total. The CS1 sample is larger because CS1 is a required course for other non-Computer Science majors and also satisfies a general University Studies requirement.
CS1 and CS2 were taught by 16 and 10 different faculty respectively in this period. 
 Earning a C or higher is required to progress from both CS1 and CS2, and students may repeat a course if they earn less than a C. Approximately 7\% of students repeat CS1 and CS2.   
 Students' grades from all course attempts are included in this section's analysis; we exclude withdrawals and audits.

\subsection{CS1 and CS2 Grade Point Analysis}

\begin{figure}[tb]
    \centering
    \includegraphics[width=\linewidth]{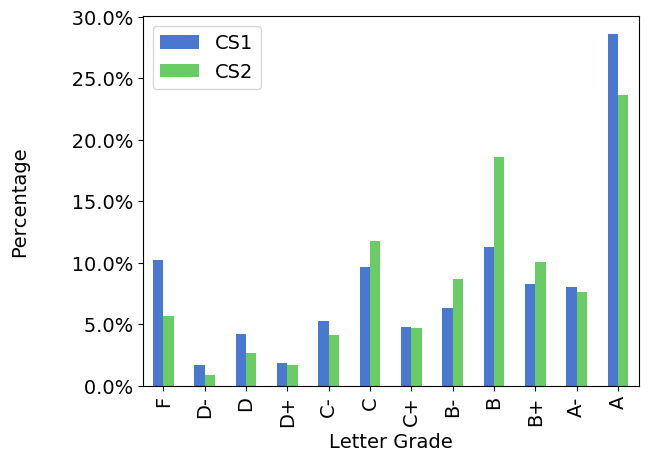}
    \caption{CS1 and CS2 Course Letter Grade Distributions, Fall 2013---Spring 2019, CS1 $n=1641$, CS2 $n=706$. Percentages Are Shown to Normalize Values between Courses}
    \label{fig:historical_distribution}
\end{figure}

 Figure~\ref{fig:historical_distribution} shows the historical distribution of grades. CS1 has a greater proportion of A and F grades, while CS2 has a greater proportion of Bs. The difference between frequency distributions of grades is statistically significant ($X^2 = 49.659, p < 0.001, df=11$). We note that grade points earned is subject to heterogeneity in grading schemas between courses, and student performance in an individual class may be attributable to total course load and out-of-class reasons. The sample size should mitigate some of the effects, but we do not have data on the size of such effects.
 
 The outright failure rate of CS1 is 10.2\% (167 Fs), whereas the failure rate in CS2 is 5.7\% (40 Fs).  The data suggests that students are more likely to excel or fail in CS1, whereas more students occupy the middle ground in CS2.  The percentage of students who do not meet the C threshold required for dependant courses are 23.1\% in CS1 and 15.0\% in CS2. Our pass rates for CS1 are higher than published large-scale studies~\cite{Bennedsen2007, Watson2014}; large studies of CS2 pass rates are not available. 
 
\begin{figure}[tb]
    \centering
    \includegraphics[width=\linewidth]{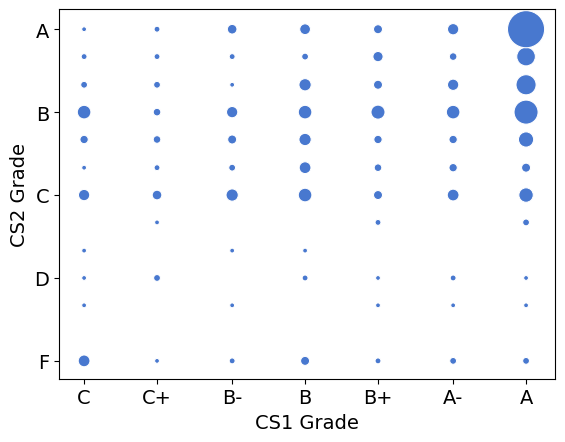}
    \caption{Scatterplot of Students' Most Recent CS1 Grades vs. CS2 Grades, $n=614$\protect\footnotemark}
    \label{fig:grade_points}
\end{figure}

We examine the relationship between a student's most recent CS2 and CS1 course grades.\footnote{We use the most recent grade, as opposed to an average grade across repeated attempts, to simplify the analysis.} The scatterplot in Figure~\ref{fig:grade_points} shows no obvious relationship between a students' CS1 and CS2 grades, though a sizeable portion of students earn As in CS1 and similarly high grades in CS2.

\begin{figure}[tb]
    \centering
    \includegraphics[width=\linewidth]{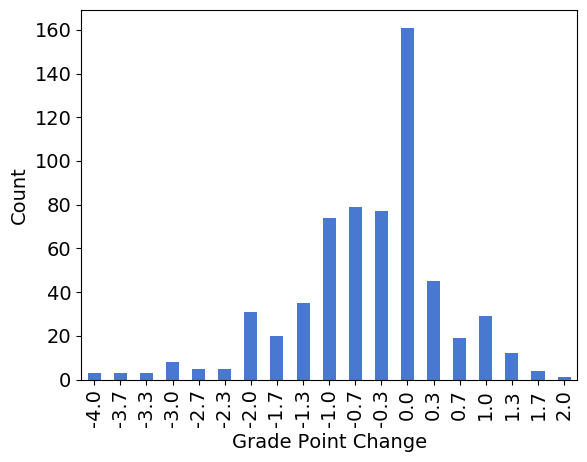}
    \caption{Grade Point Change from CS1 to CS2, $n=614$}
    \label{fig:gp_change}
\end{figure}

A more insightful analysis is to examine the \textit{grade point difference} of individuals between CS1 and CS2 as shown in Figure~\ref{fig:gp_change}. \textit{Grade point difference} is calculated by subtracting CS1 grade points earned from CS2 grade points earned on the standard 4.0 scale. All students in this sample must have passed CS1 with at least a C (with many earning As), thus the distribution is biased to be skew left. The data indicates that most student grades (56\%) dropped slightly. Only 18\% improved their letter grades, and 26\% earn the same grade bolstered by the 118 students who earned As in both courses. Figure~\ref{fig:gp_change} supports previous findings that CS1 grade is related to CS2 grade~\cite{Enbody2009, Bisgin2018} in the sense that 62.1\% of grade point changes are within one standard deviation ($\sigma=0.949$). The change distribution is not normal (D'Agostino's $K^2 = 67.29, p < 0.001$), possibly owing in part to the prerequisite grade of C in CS1.

Table~\ref{tab:pass_count} lists the letter grades earned in CS1 and the counts of students with those grades who then progress from CS2 by earning a grade of C or better. Over 97\% of students who earn an A in CS1 also earn a C or better in CS2. However, of the students who earned a C in CS1, 27\% do not earn a C or better in CS2. This amounts to only 14 students in our dataset, but they make up a disproportionate percentage of those who do not move on from CS2 ($X^2 = 38.756, p < 0.001, df=6$). 

Those who do not progress from CS2 are of particular interest as they have demonstrated at least C-level competence in CS1. 
If the proportion of students who earned a C in CS1 remains disproportionately high, then we might consider whether a grade of C in CS1 measures the necessary aptitudes for success in the program. We are also interested in those students who performed excellently in CS1 (A and A-) but failed to earn a C in CS2. We do not have insights into what drives these students' performances yet, but that is the subject of future work.

\footnotetext{The scatterplot shows students who earned grades in CS1 and CS2 at \blinduni{UNCW}; students who earned CS1 credit at other institutions are excluded which is why the number of data points (614) is less than the number of CS2 grades collected (706).}

\begin{table}[tb]
\caption{Proportions of Students Who Earned a C or Better in CS2 Grouped by CS1 Grade}
\label{tab:pass_count}
\centering
\begin{tabular}{@{}lll|r@{}}
                  & \multicolumn{2}{c}{\textit{C or better in CS2?}} &                         \\
CS1 - Final Grade & False              & True              & Total                   \\ \midrule
C                 & 14  (27\%)               & 37 (7\%)                & 51                      \\
C+                & 5   (10\%)               & 24 (4\%)               & 29                      \\
B-                & 4   (8\%)                & 41   (7\%)             & 45                      \\
B                 & 9   (17\%)                & 77  (14\%)              & 86                      \\
B+                & 6   (11.5\%)               & 51  (9\%)              & 57                      \\
A-                & 6  (11.5\%)                & 60  (11\%)              & 66                      \\
A                 & 8    (15\%)              & 272  (48\%)             & 280                     \\ \midrule
Total             & 52                 & 562               & 614
\end{tabular}
\end{table}

\section{SCS1 Score as an Indicator of CS2 Grade}
\label{sec:leading_indicators}
The second research question we investigate is: \textit{Does a quantitative relationship exist between CS2 course
grade and a modified SCS1 concept inventory score?}

Our second research objective is to evaluate a modified version (explained below) of the \textit{SCS1 concept inventory}~\cite{Parker2016} as a leading indicator of CS2 performance. The SCS1 is a validated instrument for evaluating students' understanding of select CS1 concepts.  A relationship between our modified SCS1 (mSCS1) score and CS2 final grade may indicate that the mSCS1 can be used early in the CS2 semester to identify areas of weakness.

\subsection{The SCS1 and Modifications for this Study}
\textit{Note:} We are not at liberty to disclose the SCS1 or mSCS1 per the license of the SCS1 authors, but a request for free educator access to the SCS1 can be made via its Google Group located at \url{https://goo.gl/MiYOFk}.

The Second CS1 Assessment (SCS1)~\cite{Parker2016} is derived from the Foundational CS1 (FCS1) Assesment~\cite{Tew2010}. The FCS1 was shown to have a significant, moderate correlation with CS1 Final Exam Scores~\cite{Tew2010, Tew2011} in a large, multi-institutional, multi-programming language study. The SCS1 is an "isomorphic version" of the FCS1 available to educators while the FCS1 is copyrighted by its author. Both assessments are written in pseudocode and inventory the CS1 concepts of arrays, basics, for-loops, function parameters, function return values, if-else, logical operators, recursion, and while-loops. Each of the 27 questions has five multiple choice answers, and each topic has a definition-oriented, code tracing, and code completion question. Both the FCS1 and SCS1 were intended for a 60-minute exam period. We chose the SCS1 as a candidate for leading indicators because it is language agnostic, has been validated at scale, and is freely available to educators. 

We modified the SCS1 for our study in response to community feedback on the SCS1 Google group and to fit our institutions CS1 learning outcomes and language (Python). Questions on recursion were dropped as recursion is not part of CS1 learning objectives at our institution. We translated the questions to Python as we found the pseudocode non-intuitive in places and did not wish to burden the students with interpreting the pseudolanguage. Finally, one question for each concept was dropped to encourage completion of the assessment in a 50-minute lecture session based on feedback from students and instructors in the FCS1 study~\cite{Tew2011} and on the SCS1 Google Group states that 60 minutes is not enough time to complete the original 27 questions. Questions with very low correct response rates (<20\%) reported in \cite{Tew2011} were dropped along with those that did not directly translate to Python. 

Two sections of CS2 were administered the mSCS1 assessment consisting of 16 questions translated to Python. The assessment was given closed-book in a 50-minute class session. Scrap paper was permitted. A total of 44 students were administered the test, and 39 of those completed the semester.

% \noindent\underline{\textbf{Hurricane Florence note}}: Of the 74 students who reported the Semester when they last took CS1, 59 (78.7\%) took CS1 in Fall 2019 when Hurricane Florence closed UNCW for one month. Instruction during the Fall 2019 semester was atypical in all regards. Thus, any findings and results from the Spring 2019 CS2 sample may not be representative of the population of introductory programming students.

\subsection{Course Letter Grade and mSCS1 Score}

\begin{figure}[bt]
    \centering
    \includegraphics[width=\linewidth]{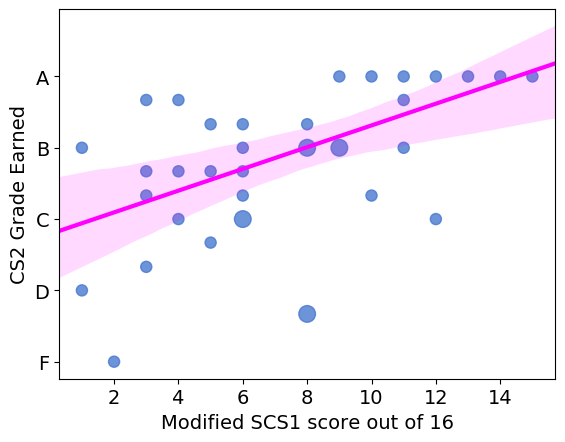}
    \caption{Modified SCS1 Score vs. CS2 Grade Points Earned, $n=39$}
    \label{fig:231_written_assessment_vs_letter_grade}
\end{figure}

Students' CS2 course grade points are plotted against their mSCS1 scores in Figure~\ref{fig:231_written_assessment_vs_letter_grade}. 
D'Agostino's statistical test does \textit{not} reject the null hypothesis that the mSCS1 scores are drawn from a normal distribution ($K^2 = 2.794, p = 0.247$). However, the distribution of CS2 final grade is not normal ($K^2 = 5.992, p = 0.049$). Spearman's non-parametric rank-order test yielded a moderate, statistically significant correlation ($\rho = 0.568, p < 0.001$). 
This suggests that the written assessment may provide a leading indicator to overall performance in the course. The sample size is relatively small (n=39), and we will replicate this analysis in future semesters. Future iterations of this work will examine student performance on the individual questions.

\section{Student Reflection Questionnaire on CS1 and CS2}
\label{sec:questionnaire}
The final research question we investigate is: \textit{What are the most challenging aspects of CS2, and how well does CS1 prepare students for CS2 from the student's perspective?}

In the final weeks of the semester, CS2 students were given a questionnaire that asked the following questions:
\begin{enumerate}
    \item Rate your agreement with the following statement: "Overall, I think that CS1 (or equivalent) adequately prepared me for this course". (1-Strongly Disagree through 5-Strongly Agree) 
    \item Which concepts or topics from CS1 were most helpful to you in this course? (free text)
    \item Which concepts or topics from CS1 would you like to have been better at prior to taking this course? (free text)
    \item What were the most challenging aspects of this course? (free text)
\end{enumerate}
Of the 79 students enrolled at the end of the semester, 69 (87\%) completed the exit questionnaire. The questionnaire was not anonymous so that student responses could be linked to course performance; that analysis will be a part of future work, and we acknowledge that the lack of anonymity may alter the truthfulness of answers.

Of the 69 respondents, 59\% agreed, 19\% were neutral, and 22\% disagreed with the statement that CS1 prepared them well for CS2. Ideally, student responses would be in the "agree" categories, but there are many personal~\cite{Bergin2005b, Pappas2016} and pedagogical reasons~\cite{Vihavainen2014, Zingaro2018} why this would not be the case irrespective of any disconnect between exit criteria from CS1 and starting expectations in CS2.

Responses to Questions 2--4 were transcribed into an Excel file and \textit{open coded}~\cite{Strauss1987} by the first author to identify topics/concepts in each response. 
A single response could mention multiple topics. Unique responses or responses that did not readily identify a topic or concern were not coded. Tables \ref{table:most_helpful}, \ref{table:would_like}, and \ref{table:most_challenging} show codes (topics and concepts) appearing in more than 10\% of the responses to each question. 

\begin{table}[b]
\caption{Questionnaire: Which Concepts or Topics from CS1 Were Most Helpful to You in this Course?}
\label{table:most_helpful}
\centering
\begin{tabularx}{\linewidth}{Xr}
Topic                                                      & Responses \\ \midrule
Foundations (basic expressions, boolean values, syntax, vocabulary) & 29 (43\%)                \\
Iteration, for-loops, or while-loops                                & 25 (37\%)                 \\
File manipulation (reading, writing, loading into a data structure) & 15 (22\%)                 \\
Recursion                                                           & 15 (22\%)                 \\
Lists and list operations                                           & 10 (15\%)                 \\
Function call and definition                                        & 10 (15\%)                \\
if-else, conditionals                                               & 8  (12\%)                
\end{tabularx}

\end{table}

Question 2 asked, "Which concepts or topics from CS1 were most helpful to you in this course?" The student responses (Table~\ref{table:most_helpful}) reflect the skills required to interact with data structures implemented as Python classes. Reading data from files, iterating over lists, and list manipulation are skills used in nearly every assignment. Data structures are implemented as Python classes with accompanying methods, e.g., a Stack class with \texttt{push()}, \texttt{pop()}, \texttt{top()}, etc. Thus, a strong grasp of function definition, call, and return is beneficial for students. Students who were exposed to recursion in CS1 found it to be useful when it is introduced in CS2. Question 3 asked, "Which concepts or topics from CS1 would you like to have been better at prior to taking this course?" The students' responses (Table~\ref{table:would_like}) again reflect the focus of CS2 on Python classes, reading files for data content, and manipulating lists. File I/O, Python lists, and function definitions are covered extensively in our CS1 curriculum, whereas recursion and class definition are not part of the CS1 learning outcomes. The responses to Questions 2 and 3 are useful for informing discussions of what \textit{could} be covered in CS1, but more importantly, these responses identify the concepts that should be reviewed, reinforced, and possibly quantified as potential leading indicators of performance at the beginning of CS2. 

\begin{table}[tb]
\caption{Questionnaire: Which Concepts or Topics from CS1 Would You Like to Have Been Better at Prior to Taking this Course?}
\label{table:would_like}
\centering
\begin{tabularx}{\linewidth}{Xr}
Topic                                                      & Responses \\ \midrule
Classes/objects                                                     & 18 (26\%)                \\
File manipulation (reading, writing, reading into a data structure) & 12 (17\%)                 \\
Lists and list operations                                           & 11 (16\%)                 \\
Recursion                                                           & 11 (16\%)                \\
Dictionaries                                                        & 8  (12\%)                
\end{tabularx}
\end{table}

\begin{table}[tb]
\caption{Questionnaire: What Were the Most Challenging Aspects of this Course?}
\label{table:most_challenging}
\centering
\begin{tabular}{lr}
Topic                   & Responses \\ \midrule
Linked list / doubly linked list & 11 (16\%)          \\
Creating code from scratch       & 10 (15\%)          \\
Classes/objects                  & 8 (12\%)           \\
Pace of instruction / work       & 7 (10\%)          
\end{tabular}

\end{table}

Question 4 asked "What were the most challenging aspects of this course? (open-ended)" concerning CS2. The responses to this question were predictably diffuse, but students did agree on some issues. Linked lists and its emphasis on pointers and class/object definition are challenging topics, reinforcing findings in Simon et al.~\cite{Simon2010} and Yeomans et al.~\cite{Yeomans2019}.
Responses discussing the Linked List and Class/Objects topics often reference challenges in translating the concepts into code. That is, students could understand the notion of Linked Lists and Objects but had difficulty implementing the data structures and applying them to problems. This likely relates to students' mental models of how the computer processes programs and points to a particular area where pedagogy can be improved. The "Creating Code from Scratch" topic refers to responses where students remarked on how past instructors required them to complete a half-finished Python script, but now they were required to create Python scripts from scratch or using a minimal code template. CS2 instructors should be aware of this paradigm. More importantly, student grades (particularly in CS2) may be predicated on their ability to decompose problems and initiate a solution with more independence than was required in the past. Thus, instructors may need to include these skills in course content and explicitly add them to course learning outcomes and assessments.

\section{Threats to Validity}
We discuss this study's threats to validity according to the categories of Cook and Campbell~\cite{cook1979}.

\subsection{Conclusion Validity}
Conclusion validity refers to issues that affect the ability to associate treatment and outcome. \textit{Low statistical power} is not a concern for our CS1-to-CS2 grade point comparison as the sample sizes are sufficiently large, however, the correlation between mSCS1 score and CS2 Grade Points Earned (Figure~\ref{fig:231_written_assessment_vs_letter_grade}) had a sample size of 39, which is relatively low. Appropriate non-parametric statistical tests were used in all analyses as none of the data were normally distributed. The \textit{reliability of the mSCS1} assessment has not been evaluated, and thus we cannot be certain if the outcomes of that assessment are replicable across contexts, especially outside of the Python language. Further, students' answers are likely influenced by the specific pedagogical style of the instructor, course content, and assignment content. The coding performed in this section was only performed by the first author due to time constraints, and thus interrater reliability is not available. We will continue administering the questionnaire in future semesters to improve the reliability of those responses.

\subsection{Internal Validity}
Internal validity concerns surround the causal inference between treatment and outcome should a relationship be detected. Some students took CS2 multiple times, introducing a potential \textit{history effect} wherein their questionnaire answers and classroom performance was perhaps different than those of students enrolled only one time. We do not isolate these students in our analyses and cannot quantify any such effects. Table~\ref{tab:repeat_counts} shows the number of repeats for each course; approximately 6\% and 7\% of students repeat CS1 and CS2 respectively. Only students' most recent grades are included in the calculation of Grade Change between CS1 and CS2 (Figure~\ref{fig:gp_change}). Finally, our questionnaire in Section~\ref{sec:questionnaire} was not anonymous, and thus students may not have been compelled to answer truthfully. 

\begin{table}[htb]

\caption{Individuals' Times Taken Excluding Withdrawals and Audits}
\centering
\begin{tabular}{@{}rrr@{}}
Times Taken & CS1 Count & CS2 Count \\ \midrule
1           & 1434       & 609       \\
2           & 91        & 44        \\
3           & 7         & 3         \\
4           & 1         & -         \\ \midrule
Total       & 1533       & 656    
\end{tabular}
\label{tab:repeat_counts}
\end{table}

\subsection{Construct Validity}
Construct validity concerns issues around whether the analysis results are connected to the driving theory, in our case, whether we adequately captured issues related to success in CS2 and the CS1-CS2 transition. One goal of our study (hopefully the first in a series) was to help better operationalize these constructs. Additional methods, such as semi-structured interviews and analyses of individual homework assignments and exam questions, will be used in the future to better capture which topics from CS2 are most challenging for students. Further, CS2 Grade is only one indicator of CS2 performance. In the future, we will also use the validated Basic Data Structures Inventory (BDSI)~\cite{Porter2019} to assess student understanding of CS2 concepts.

\subsection{External Validity}
External validity concerns the transferability of results outside the study context. We describe the content and environment of our CS1 and CS2 courses in Section~\ref{sec:curriculum}, and it is up to the reader to decide if our Python-based courses resemble their own. One point of note is that our failure rate for CS1 is lower than that reported in other large scale studies, and thus our average CS1 grades may be inflated. No such data is available for CS2. 

\subsection{Conclusions and Future Work}
\label{sec:conclusion}
Our long term research goal is to investigate \textit{successful and unsuccessful transitions from CS1 to CS2}. As the first phase of that research, we provide a quantitative analysis of CS1 and CS2 grade point statistics--an evaluation of a modified version of the SCS1 concept inventory as an indicator of CS2 performance--and performed qualitative analysis of a questionnaire that obtained feedback from CS2 students on helpful CS1 preparation and challenges in CS2. In a sample of 614 students, we find that CS1 grades appear to have a statistical relationship with CS2 grades. This corroborates earlier findings~\cite{Enbody2009, Bisgin2018}, but we provide a much larger sample size. \textit{Further, the majority (56\%) of students' CS2 grades were lower than their CS1 grade, though 62\% of the grade point changes were within one standard deviation}. A disproportionate number of C students from CS1 do not pass CS2 owing to the mean grade drop. 

Our goal as educators is to ensure that students exiting CS1 have the necessary capabilities to succeed in CS2. To this end, we administered a modified SCS1 concept inventory~\cite{Parker2016} and correlated the scores with CS2 grade points earned. \textit{We found a statistically significant correlation between mSCS1 score and CS2 grade points}, suggesting that the concepts on the mSCS1 and/or the way the mSCS1 tests those concepts measure useful information about beneficial skills in CS2. We believe, anecdotally, that strong performance on the mSCS1 requires solid mental models of \textit{how} programs are interpreted by the machine. We believe that these mental models are central to both program debugging and abstraction. We will continue to investigate the individual questions and concepts captured in the mSCS1 for continued predictive power of CS2 grade and to identify possible improvements in CS1 structure.

From our questionnaire, the majority of students believed that CS1 adequately prepared them for CS2. The only data structure explicitly mentioned as challenging was \textit{linked lists}. This may be a consequence of using Python as the programming language of choice. Unlike C/C++, Python (and Java) does not make the concept of pointers or memory references explicit. \textit{Our findings are in line with studies have identified pointers and classes as threshold concepts}~\cite{Sanders2012, Yeomans2019} for students learning to program. Students indicated that coding a data structure to solve a problem, rather than filling in missing parts of an implementation, was challenging. This finding is congruous with previous papers who found that of \textit{data structure application is a challenging addition to the finer points of data structure implementation}~\cite{Simon2010, Porter2018}.

A tremendous amount of computer science education research exists on success and failure factors in CS1, and a growing set of literature is being produced for CS2. We will continue to investigate cross-course indicators of success, specifically through more thorough investigations of the SCS1 inventory, the new Basic Data Structures Inventory~\cite{Porter2019}, in-process code execution and class participation metrics (e.g.,~\cite{Carter2017}), and qualitative feedback from the students themselves. We hope that we can improve students outcomes in CS2 (and beyond) through interventions driven by quantitative observation, and ensure that those students who commit to CS2 as part of a computing path receive the best opportunity to gain the skills and knowledge for career success.

\begin{acks}
Thanks to the students of \blind{CSC231} who contributed feedback to this study. Thanks also to \blind{Elham Ebrahimi}, \blind{Clayton Ferner}, \blind{Sridhar Narayan}, \blind{Toni Pence}, \blind{Geoffrey Stoker} and other CS1 and CS2 instructors for their valuable participation and input in this study.
\end{acks}

\balance
\bibliographystyle{ACM-Reference-Format_abbrev}
\bibliography{references}

\appendix

\end{document}